\documentclass[fleqn,11pt]{wlscirep}
\usepackage{epstopdf}
\title{Accelerating Adaptive IDW Interpolation Algorithm on a Single GPU}

\author[1,*]{Gang Mei}
\author[1]{Liangliang Xu}
\author[1,*]{Nengxiong Xu}
\affil[1]{School of Engineering and Technology, China University of Geosciences, Beijing, China}

\affil[*]{corresponding: \{gang.mei; xunengxiong\}@cugb.edu.cn}


\begin{abstract}
This paper focuses on the design and implementing of GPU-accelerated 
Adaptive Inverse Distance Weighting (AIDW) interpolation algorithm. The AIDW 
is an improved version of the standard IDW, which can adaptively determine 
the power parameter according to the spatial points' distribution pattern 
and achieve more accurate predictions than those by IDW. In this paper, we 
first present two versions of the GPU accelerated AIDW, the naive version 
without profiting from shared memory and the tiled version taking advantage 
of shared memory. We also implement the naive version and the tiled version 
using the data layouts, Structure of Arrays (AoS) and Array of aligned 
Structures (AoaS), on single and double precision. We then evaluate the 
performance of the GPU-accelerated AIDW by comparing it with its original 
CPU version. Experimental results show that: on single precision the naive 
version and the tiled version can achieve the speedups of approximately 270 
and 400, respectively. In addition, on single precision the implementations 
using the layout SoA are always slightly faster than those using layout 
AoaS. However, on double precision, the speedup is only about 8; and we have 
also observed that: (1) there are no performance gains obtained from the 
tiled version against the naive version; and (2) the use of SoA and AoaS 
does not lead to significant differences in computational efficiency. 
\end{abstract}

\begin{document}

\flushbottom
\maketitle
%
%
\thispagestyle{empty}

\section{Introduction}
A spatial interpolation algorithm is the method in which the attributes at 
some known locations (data points) are used to predict the attributes at 
some unknown locations (interpolated points). Spatial interpolation 
algorithms such as the Inverse Distance Weighting (IDW) \cite{01}, Kriging \cite{02}, and Discrete Smooth Interpolation (DSI) \cite{03,04}, are 
commonly used in geosciences and related research fields, especially in 
Geographic Information System (GIS); see a brief summary in \cite{05} and 
a comparative survey in \cite{06}. 

Among the above mentioned three spatial interpolation algorithms, only the 
Kriging method is computationally intensive due to the inversion of the 
coefficient matrix, while the other two are easy to compute. However, when 
the above three algorithms are applied to a large set of points, for example 
more than 1 million points, they are still quite computationally expensive, 
even for the simplest interpolation algorithm IDW. 

To be able to apply those interpolation algorithms in practical 
applications, it is needed to improve the computational efficiency. With the 
rapid development of multi-core CPU and GPU hardware architecture, parallel 
computing technology has made remarkable progress. One of the most effective 
and commonly used strategies for enhancing the computational efficiency of 
interpolation algorithms is to parallelize the interpolating procedure in 
various massively parallel computing environments on multicore CPUs and/or 
GPUs platforms. 

Starting in the 1990s, many researchers have devoted themselves to the 
parallelization of various interpolation algorithms \cite{07,08,09,10,11}. Specifically 
for the Kriging method, many parallel programs were implemented on high 
performance and distributed architectures \cite{10,12,13,14,15,16,17,18,19,20}. Also, to reduce 
the computational cost in large-scale applications, the IDW algorithm has 
been parallelized in various massively parallel computing environments on 
multicore CPUs and/or GPUs platforms.

For example, by taking advantage of the power of traditional CPU-based 
parallel programming models, Armstrong and Marciano \cite{07,08} 
implemented the IDW interpolation algorithm in parallel using FORTRAN 77 on 
shared-memory parallel supercomputers, and achieved an efficiency close to 
0.9. Guan and Wu \cite{09} performed their parallel IDW algorithms 
used open multi-processing (OpenMP) running on an Intel Xeon 5310, achieving 
an excellent efficiency of 0.92. Huang, Liu, Tan, Wang, Chen 
and He \cite{21} designed a parallel IDW interpolation algorithm with the message 
passing interface (MPI) by incothe message passing interfacecess, multiple 
data (SPMD) and master/slave (M/S) programming modes, and attained a speedup 
factor of almost 6 and an efficiency greater than 0.93 under a Linux cluster 
linked with six independent PCs. Li, Losser, Yorke and Piltner \cite{22} developed the parallel 
version of the IDW interpolation using the Java Virtual Machine (JVM) for 
the multi-threading functionality, and then applied it to predict the 
distribution of daily fine particulate matter PM 2.5.

Since that general purpose computing on modern Graphics Processor Units 
(GPUs) can significantly reduce computational times by performing massively 
parallel computing, current research efforts are being devoted to parallel 
IDW algorithms on GPU computing architectures such as CUDA \cite{23}
 and OpenCL \cite{24}. For example, Huraj, Sil\'{a}di and Sil\'{a}\u{c}i \cite{25,26}
have deployed IDW on GPUs to accelerate snow cover depth prediction. 
Henneb\"{o}hl, Appel and Pebesma \cite{13} studied the behavior of IDW on a 
single GPU depending on the number of data values, the number of prediction 
locations, and different ratios of data size and prediction locations. 
Hanzer \cite{27} implemented the standard IDW algorithm using Thrust, PGI 
Accelerator and OpenCL. Xia, Kuang and Li \cite{28,29} developed the GPU 
implementations of an optimized IDW algorithm proposed by them, and obtained 
13--33-fold speedups in computation time over the sequential version.

And quite recently, Mei \cite{30} developed two GPU implementations of 
the IDW interpolation algorithm, the tiled version and the CDP version, by 
taking advantage of shared memory and CUDA Dynamic Parallelism, and found 
that the tiled version has the speedups of 120 and 670 over the CPU 
version when the power parameter $p$ was set to 2 and 3.0, respectively, but the 
CDP version is 4.8 $\sim $ 6.0 times slower than the naive GPU version. In 
addition, Mei \cite{31} compared and analyzed the impact of data layouts 
on the efficiency of GPU-accelerated IDW implementations.

The Adaptive Inverse Distance Weighting (AIDW) interpolation algorithm \cite{32} is an improved version of the standard IDW. The standard IDW is relatively fast and easy to compute, and straightforward to 
interpret. However, in the standard IDW the distance-decay parameter is 
applied uniformly throughout the entire study area without considering the 
distribution of data within it, which leads to less accurate predictions 
when compared other interpolation methods such as Kriging \cite{32}. In the AIDW, the distance-decay parameter is a no longer 
constant value over the entire interpolation space, but can be adaptively 
calculated using a function derived from the point pattern of the 
neighborhood.

The AIDW performs better than the constant parameter method in most cases, 
and better than ordinary Kriging in the cases when the spatial structure in 
the data could not be modeled effectively by typical variogram functions. In 
short, the standard IDW is a logical alternative to Kriging, but AIDW offers 
a better alternative. 

As stated above, when exploited in large-scale applications, the standard 
IDW is in general computationally expensive. As an improved and complicated 
version of the standard IDW, the AIDW in this case will be also 
computationally expensive. To the best of the authors' knowledge, however, 
there is currently no existing literature reporting the development of 
parallel AIDW algorithms on the GPU.

In this paper, we introduce our efforts dedicated to designing and 
implementing the parallel AIDW interpolation algorithm \cite{32} on 
a single modern Graphics Processing Unit (GPU). We first present a 
straightforward but suitable-for-paralleling method for finding the nearest 
points. We then develop two versions of the GPU implementations, i.e., The 
naive version that does not take advantage of the shared memory and the 
title version that profile from the shared memory. We also implement both 
the naive version and the tiled version using two data layouts to compare 
the efficiency. We observe that our GPU implementations can achieve 
satisfied speedups over the corresponding CPU implementation for varied 
sizes of testing data.

Our contributions in this work can be summarized as follows:

(1) We present the GPU-accelerated AIDW interpolation algorithm;

(2) We provide practical GPU implementations of the AIDW algorithm.


The rest of this paper is organized as follows. Section \ref{sec2} gives a brief 
introduction to the AIDW interpolation. Section \ref{sec3} introduces considerations 
and strategies for accelerating the AIDW interpolation and details of the 
GPU. Section \ref{sec4} presents some experimental tests that are performed on single 
and/or double precision. And Section \ref{sec5} discusses the experimental results. 
Finally, Section \ref{sec6} draws some conclusions.

\section{IDW and AIDW Interpolation }\label{sec2}
\subsection{The Standard IDW Interpolation}
The IDW algorithm is one of the most commonly used spatial interpolation 
methods in Geosciences, which calculates the prediction values of unknown 
points (interpolated points) by weighting average of the values of known 
points (data points). The name given to this type of methods was motivated 
by the weighted average applied since it resorts to the inverse of the 
distance to each known point when calculating the weights. The difference 
between different forms of IDW interpolation is that they calculate the 
weights variously. 

A general form of predicting an interpolated value $Z$ at a given point $x$ based 
on samples $Z_{i}=Z(x_{i})$ for $i$ = 1, 2, {\ldots}, $n$ using IDW is an 
interpolating function: 
\begin{equation}
\label{eq1}
Z(x)={\sum\limits_{i=1}^n }{\frac{\omega _i (x)z_i }{\sum\limits_{j=1}^n 
		{\omega _j (x)} }} ,
\quad
\omega _i (x)=\frac{1}{d(x,x_i )^\alpha }.
\end{equation}

The above equation is a simple IDW weighting function, as defined by Shepard \cite{01}, where $x$ denotes a predication location, $x_{i }$ is a data 
point, $d$ is the distance from the known data point $x_{i}$ to the unknown 
interpolated point $x$, $n$ is the total number of data points used in 
interpolating, and $p$ is an arbitrary positive real number called the power 
parameter or the distance-decay parameter (typically, $\alpha $ = 2 in the 
standard IDW). Note that in the standard IDW, the power/distance-decay 
parameter $\alpha $ is a user-specified constant value for all unknown 
interpolated points. 

\subsection{The AIDW Interpolation}
The AIDW is an improved version of the standard IDW, which is originated by 
Lu and Wong \cite{32}. The basic and most important idea 
behind the AIDW is that: it adaptively determines the distance-decay 
parameter $\alpha $ according to the spatial pattern of data points in the 
neighborhood of the interpolated points. In other words, the distance-decay 
parameter $\alpha $ is no longer a pre-specified constant value but 
adaptively adjusted for a specific unknown interpolated point according to 
the distribution of the data points/sampled locations.

When predicting the desired values for the interpolated points using AIDW, 
there are typically two phases: the first one is to adaptively determine the 
parameter $\alpha $ according to the spatial pattern of data points; and the 
second is to perform the weighting average of the values of data points. The 
second phase is the same as that in the standard IDW; see Equation (\ref{eq1}).

In AIDW, for each interpolated point, the adaptive determination of the 
parameter $\alpha $ can be carried out in the following steps.

\textbf{Step 1}: Determine the spatial pattern by comparing the observed 
average nearest neighbor distance with the expected nearest neighbor 
distance.

\begin{enumerate}[label=\arabic{*})]
	\item Calculate the expected nearest neighbor distance $r_{\exp } $ for a random pattern using:
\begin{equation}
\label{eq2}
r_{\exp } =\frac{1}{2\sqrt {n \mathord{\left/ {\vphantom {n A}} \right. 
			\kern-\nulldelimiterspace} A} },
\end{equation}
where $n$ is the number of points in the study area, and $A$ is the area of the 
study region.

	\item Calculate the observed average nearest neighbor distance $r_{obs} $ by taking the average of the nearest neighbor distances for all points:
\begin{equation}
\label{eq3}
r_{obs} =\frac{1}{k}\sum\limits_{i=1}^k {d_i } ,
\end{equation}
where $k$ is the number of nearest neighbor points, and $d_i $ is the 
nearest neighbor distances. The $k$ can be specified before interpolating.

	\item Obtain the nearest neighbor statistic $R\left( {S_0 } \right)$ by:
\begin{equation}
\label{eq4}
R\left( {S_0 } \right)=\frac{r_{obs} }{r_{\exp } },
\end{equation}
where $S_{0 }$ is the location of an unknown interpolated point.
\end{enumerate}

\textbf{Step 2}: Normalize the $R\left( {S_0 } \right)$ measure to $\mu _R $ 
such that $\mu _R $ is bounded by 0 and 1 by a fuzzy membership function: 
\begin{equation}
\label{eq5}
\mu _R =\left\{ {\begin{array}{ll}
	0&R\left( {S_0 } \right)\le \mbox{ }R_{\min } \mbox{ } \\ 
	0.5-0.5\cos \left[ {\frac{\pi }{R_{\max } }\left( {R\left( {S_0 } 
			\right)-R_{\min } } \right)} \right]&R_{\min } \le R\left( {S_0 } 
	\right)\le \mbox{ }R_{\max } \mbox{ } \\ 
	1&R\left( {S_0 } \right)\ge \mbox{ }R_{\max } \\ 
	\end{array}} \right.,
\end{equation}
where $R_{\min } \mbox{ }$ or $R_{\max } $ refers to a local nearest neighbor 
statistic value (in general, the $R_{\min } \mbox{ }$ and $R_{\max } $ can 
be set to 0.0 and 2.0, respectively).

\textbf{Step 3}: Determine the distance-decay parameter $\alpha $ by mapping 
the $\mu _{R}$ value to a range of $\alpha _{ }$ by a triangular 
membership function that belongs to certain levels or categories of 
distance-decay value; see Equation (\ref{eq6}).
\begin{equation}
\label{eq6}
\alpha \left( {\mu _R } \right)=\left\{ {{\begin{array}{ll}
		{\alpha _1 } & {\mbox{0.0}\le \mu _R \le \mbox{0.1}} \\
		{\alpha _1 \left[ {1-5\left( {\mu _R -0\mbox{.}1} \right)} \right]+5\alpha 
			_2 \left( {\mu _R -0\mbox{.}1} \right)} & {\mbox{0.1}\le \mu _R \le 
			\mbox{0.3}} \\
		{5\alpha _3 \left( {\mu _R -0\mbox{.}3} \right)+\alpha _2 \left[ {1-5\left( 
				{\mu _R -0\mbox{.}3} \right)} \right]} & {\mbox{0.3}\le \mu _R \le 
			0\mbox{.}5} \\
		{\alpha _3 \left[ {1-5\left( {\mu _R -0\mbox{.5}} \right)} \right]+5\alpha 
			_4 \left( {\mu _R -0\mbox{.}5} \right)} & {\mbox{0.5}\le \mu _R \le 
			\mbox{0.7}} \\
		{5\alpha _5 \left( {\mu _R -0\mbox{.7}} \right)+\alpha _4 \left[ {1-5\left( 
				{\mu _R -0\mbox{.7}} \right)} \right]} & {\mbox{0.7}\le \mu _R \le 
			\mbox{0.9}} \\
		{\alpha _5 } & {\mbox{0.9}\le \mu _R \le \mbox{1.0}} \\
		\end{array} }} \right.,
\end{equation}
where the $\alpha _{1}$, $\alpha _{2}$, $\alpha _{3}$, $\alpha 
_{4}$, $\alpha _{5}$ are the assigned to be five levels or categories of 
distance-decay value.

After adaptively determining the parameter $\alpha $, the desired prediction 
value for each interpolated point can be obtained via the weighting average. 
This phase is the same as that in the standard IDW; see Equation (\ref{eq1}).

\section{GPU-accelerated AIDW Interpolation Algorithm}
\label{sec3}
\subsection{Strategies and Considerations for GPU Acceleration}

\subsubsection{Overall Considerations}

The AIDW algorithm is inherently suitable to be parallelized on GPU 
architecture. This is because that: in AIDW, the desired prediction value 
for each interpolated point can be calculated independently, which means 
that it is naturally to calculate the prediction values for many 
interpolated points concurrently without any data dependencies between the 
interpolating procedures for any pair of the interpolated points.

Due to the inherent feature of the AIDW interpolation algorithm, it is 
allowed a single thread to calculate the interpolation value for an 
interpolated point. For example, assuming there are $n$ interpolation points 
that are needed to be predicted their values such as elevations, and then it 
is needed to allocated $n$ threads to concurrently calculate the desired 
predication values for all those $n$ interpolated points. Therefore, the AIDW 
method is quite suitable to be parallelized on GPU architecture.

In GPU computing, shared memory is expected to be much faster than global 
memory; thus, any opportunity to replace global memory access by shared 
memory access should therefore be exploited \cite{06}. A common optimization 
strategy is called ``tiling'', which partitions the data stored in global 
memory into subsets called tiles so that each tile fits into the shared 
memory \cite{14}.

This optimization strategy ``tiling'' is also adopted to accelerate the AIDW 
interpolation algorithm: the coordinates of data points are first 
transferred from global memory to shared memory; then each thread within a 
thread block can access the coordinates stored in shared memory 
concurrently. Since the shared memory residing in the GPU is limited per SM 
(Stream Multiprocessor), the data in global memory, that is, the coordinates 
of data points, needs to be first split/tiled into small pieces and then 
transferred to the shared memory. By employing the ``tiling'' strategy, the 
global memory accesses can be significantly reduced; and thus the overall 
computational efficiency is expected to be improved.

\subsubsection{Method for Finding the Nearest Data Points}

The essential difference between the AIDW algorithm and the standard IDW 
algorithm is that: in the standard IDW the parameter power $\alpha $ is 
specified to a constant value (e.g., 2 or 3.0) for all the interpolation 
points, while in contrast in the AIDW the power $\alpha $ is adaptively 
determined according to the distribution of the interpolated points and data 
points. In short, in IDW the power $\alpha $ is user-specified and constant 
before interpolating; but in AIDW the power $\alpha $ is no longer 
user-specified or constant but adaptively determined in the interpolating. 

The main steps of adaptive determining the power $\alpha $ in the AIDW have 
been listed in subsection 2.2. Among these steps, the most computationally 
intensive step is to find the $k$ nearest neighbors (kNN) for each interpolated 
point. Several effective kNN algorithms have been developed by region 
partitioning using various data structures \cite{20,33,34,35}. However, these 
algorithms are computationally complex in practice, and are not suitable to 
be used in implementing AIDW. This is because that in AIDW the kNN search 
has to be executed within a single CUDA thread rather than a thread block or 
grid. 

\begin{figure}[ht]
	\centering
	\includegraphics[width=0.75\linewidth]{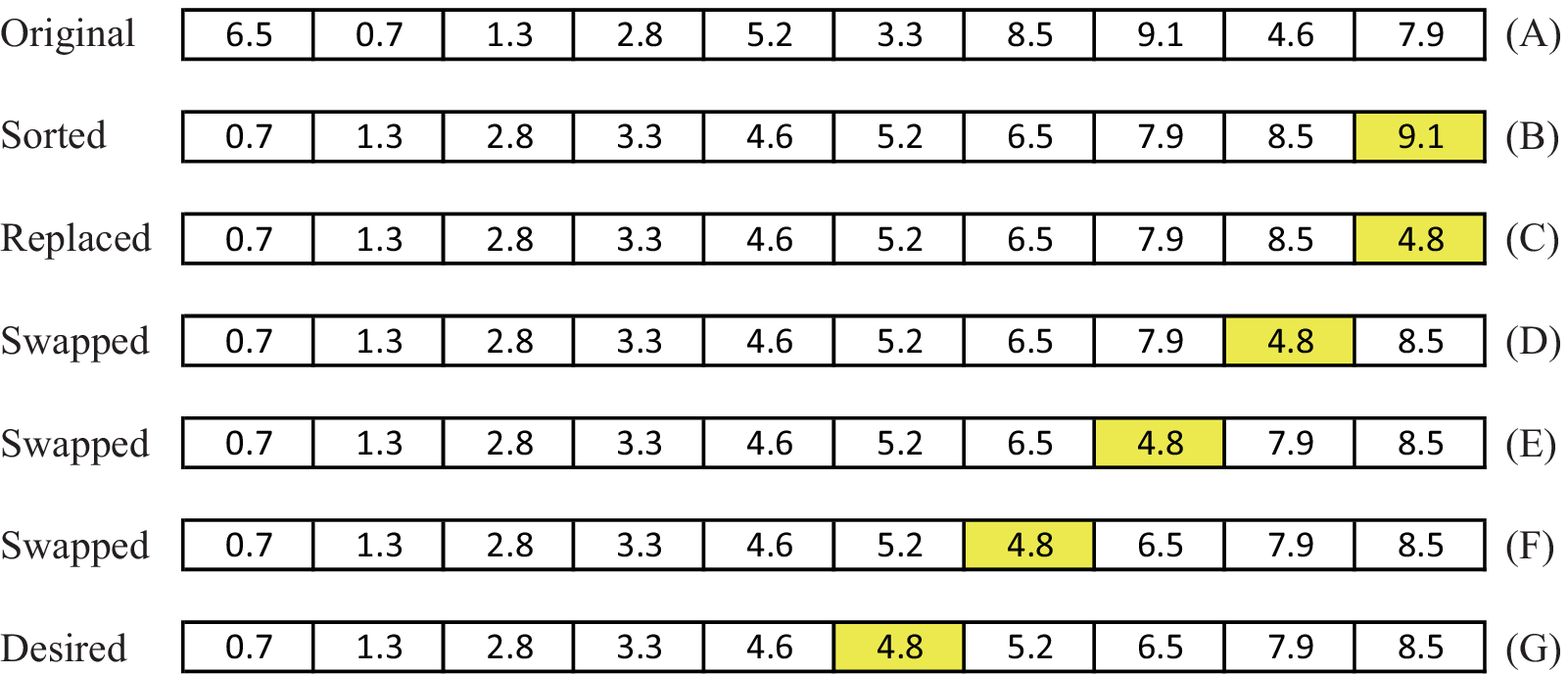}
	\caption{Demonstration of the finding of $k$ nearest neighbors ($k$ = 10)}
	\label{fig1:knn}
\end{figure}

In this paper, we present a straightforward but suitable for the GPU 
parallelized algorithm to find the $k$ nearest data points for each 
interpolated point. Assuming there are $n$ interpolated points and $m$ data 
points, for each interpolated point we carry out the following steps:

\textbf{Step 1:} Calculate the first $k$ distances between the first $k$ data points and 
the interpolated points; for example, if the $k$ is set to 10, then there are 
10 distances needed to be calculated; see the row (A) in Figure \ref{fig1:knn}.

\textbf{Step 2:} Sort the first $k$ distances in ascending order; see the row (B) in 
Figure \ref{fig1:knn}.

\textbf{Step 3:} For each of the rest ($m-k)$ data points, 

1) Calculate the distance \textit{dist}, for example, the distance is 4.8 (\textit{dist} = 4.8);

2) Compare the \textit{dist} with the $k$th distance:

if \textit{dist} $<$ the $k$th distance, then replace the $k$th distance with the \textit{dist} (see row 
(C))

3) Iteratively compare and swap the neighboring two distances from the $k$th 
distance to the 1\textsuperscript{st} distance until all the $k$ distances are newly sorted 
in ascending order; see the rows (C) $\sim $ (G) in Figure \ref{fig1:knn}.

\subsubsection{The Use of Different Data Layouts}

Data layout is the form in which data should be organized and accessed in 
memory when operating on multi-valued data such as sets of 3D points. The 
selecting of appropriate data layout is a crucial issue in the development 
of GPU accelerated applications. The efficiency performance of the same GPU 
application may drastically differ due to the use of different types of data 
layout.

\begin{figure}[ht]
	\centering
	\includegraphics[width=0.65\linewidth]{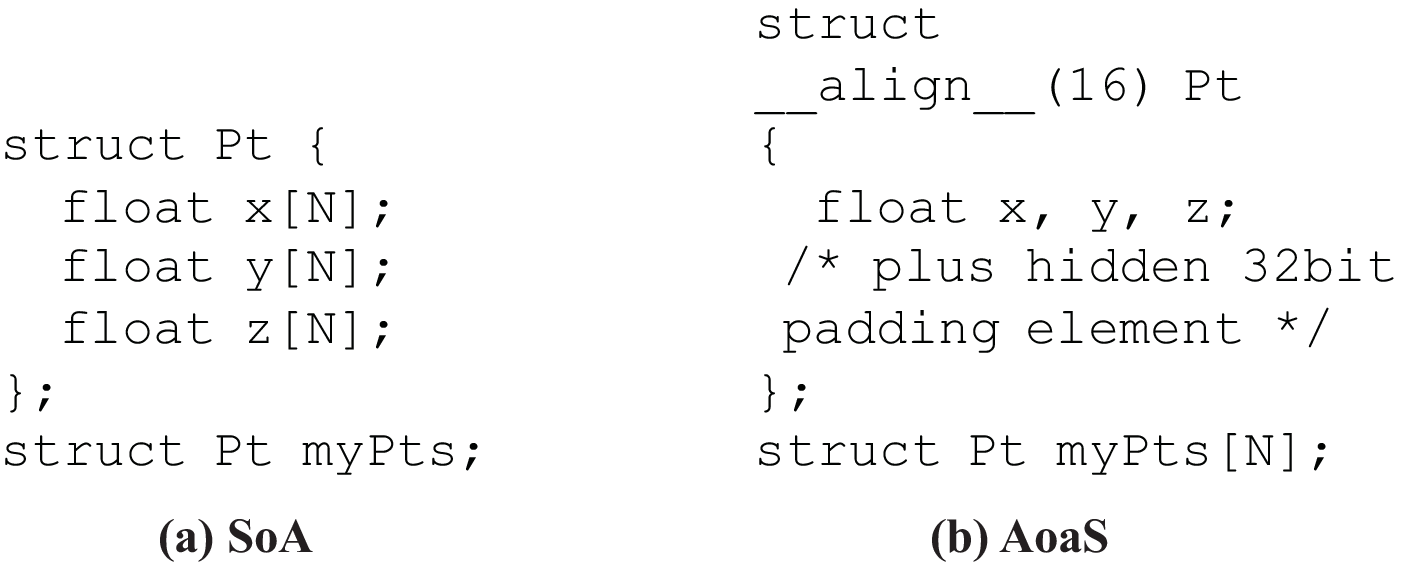}
	\caption{Data Layouts SoA and AoaS}
	\label{fig2:data layout}
\end{figure}

Typically, there are two major choices of the data layout: the Array of 
Structures (AoS) and the Structure of Arrays (SoA) \cite{36}; see Figure 
\ref{fig2:data layout}. Organizing data in AoS layout leads to coalescing issues as the data are 
interleaved. In contrast, the organizing of data according to the SoA layout 
can generally make full use of the memory bandwidth due to no data 
interleaving. In addition, global memory accesses based upon the SoA layout 
are always coalesced. 

In practice, it is not always obvious which data layout will achieve better 
performance for a specific GPU application. A common solution is to 
implement a specific application using above two layouts separately and then 
compare the performance. In this work, we will evaluate the performance 
impact of the above two basic data layouts and other layouts that are 
derived from the above two layouts. 

\subsection{Implementation Details}

This subsection will present the details on implementing the GPU-accelerated 
AIDW interpolation algorithm. We have developed two versions: (1) the 
\textit{naive} version that does not take advantage of the shared memory, and (2) the 
\textit{tiled} version that exploits the use of shared memory. And for both of the above 
two versions, two implementations are separately developed according to the 
two data layouts SoA and AoaS.

\subsubsection{Naive Version}

In this naive version, only registers and global memory are used without 
profiting from the use of shared memory. The input data and the output data, 
i.e., the coordinates of the data points and the interpolated points, are 
stored in the global memory. 

Assuming that there are $m$ data points used to evaluate the interpolated 
values for $n$ prediction points, we allocate $n$ threads to perform the 
parallelization. In other words, each thread within a grid is responsible to 
predict the desired interpolation value of one interpolated point.

\begin{figure}[htbp]
	\centering
	\includegraphics[width=0.9\linewidth]{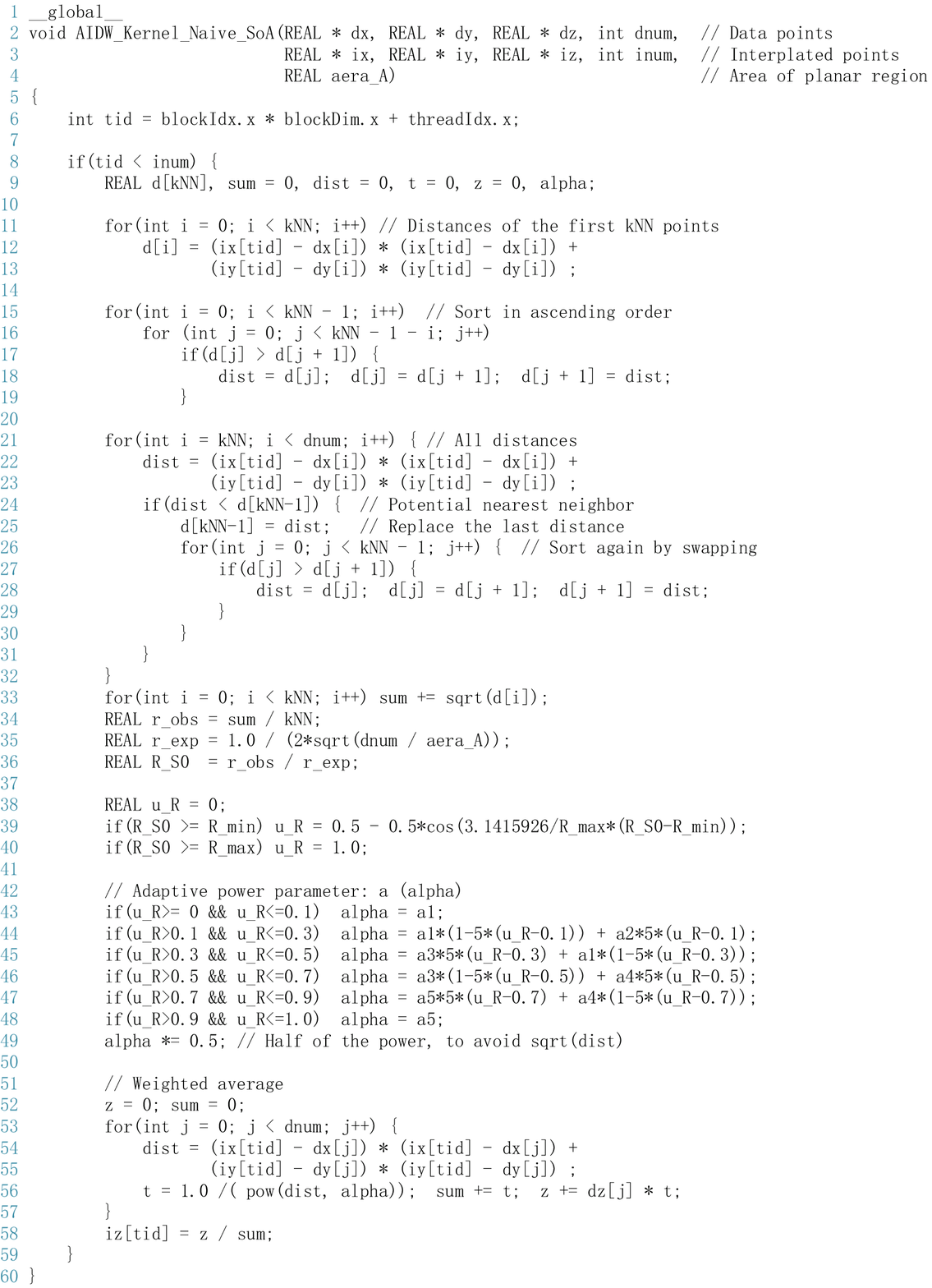}
	\caption{A CUDA kernel of the naive version of GPU-accelerated AIDW}
	\label{fig3:CUDA Kernel}
\end{figure}

A complete CUDA kernel is listed in Figure \ref{fig3:CUDA Kernel}. The coordinates of all data 
points and prediction points are stored in the arrays \texttt{REAL} 
\texttt{dx[dnum]}, \texttt{dy[dnum]}, \texttt{dz[dnum]}, \texttt{ix[inum]},\texttt{iy[inum]}, and \texttt{iz[inum]}. The word \texttt{REAL} is 
defined as float and double on single and double precision, respectively.

Within each thread, we first find the $k$ nearest data points to calculate the 
$r_{obs} $ (see Equation (\ref{eq3})) according to the straightforward approach 
introduced in subsection 3.1.2 Method for Finding the Nearest Data 
Points; see the piece of code from line 11 to line 34 in Figure \ref{fig3:CUDA Kernel}; 
then we compute the $r_{\exp } $ and $R\left( {S_0 } \right)$ according to 
Equations (\ref{eq2}) and (\ref{eq4}). After that, we normalize the $R\left( {S_0 } 
\right)$ measure to $\mu _R $ such that $\mu _R $ is bounded by 0 and 1 by a 
fuzzy membership function; see Equation (\ref{eq5}) and the code from line 38 to 
line 40 in Figure \ref{fig3:CUDA Kernel}. Finally, we determine the distance-decay parameter 
$\alpha $ by mapping the $\mu _{R}$ values to a range of $\alpha _{ }$ by 
a triangular membership function; see Equation (\ref{eq6}) and the code from line 42 
to line 49.

After adaptively determining the power parameter$\alpha $, we calculate the 
distances to all the data points again; and then according to the distances 
and the determined power parameter $\alpha $, all the $m$ weights are obtained; 
finally, the desired interpolation value is achieved via the weighting 
average. This phase of calculating the weighting average is the same as that 
in the standard IDW method.

Note that, in the naive version, it is needed to compute the distances from 
all data points to each prediction point \textit{twice}. The first time is carried out to 
find the $k $ nearest neighbors/data points, see the code from line 11 to line 
32; and the second is to calculate the distance-inverse weights; see the 
code from line 52 to line 57. 

\subsubsection{Tiled Version}

The workflow of this tiled version is the same as that of the naive version. 
The major difference between the two versions is that: in this version, the 
shared memory is exploited to improve the computational efficiency. The 
basic ideas behind this tiled version are as follows.

The CUDA kernel presented in Figure \ref{fig3:CUDA Kernel} is a straightforward implementation of 
the AIDW algorithm that does not take advantage of shared memory. Each 
thread needs to read the coordinates of all data points from global memory. 
Thus, the coordinates of all data points are needed to be read $n$ times, where 
$n$ is the number of interpolated points.

In GPU computing, a quite commonly used optimization strategy is the 
``tiling,'' which partitions the data stored in global memory into subsets 
called tiles so that each tile fits into the shared memory \cite{14}. This 
optimization strategy ``tiling'' is adopted to accelerate the AIDW 
interpolation: the coordinates of data points are first transferred from 
global memory to shared memory; then each thread within a thread block can 
access the coordinates stored in shared memory concurrently. 

In the tiled version, the tile size is directly set as the same as the block 
size (i.e., the number of threads per block). Each thread within a thread 
block takes the responsibilities to loading the coordinates of one data 
point from global memory to shared memory and then computing the distances 
and inverse weights to those data points stored in current shared memory. 
After all threads within a block finished computing these partial distances 
and weights, the next piece of data in global memory is loaded into shared 
memory and used to calculate current wave of partial distances and weights.

It should be noted that: in the tiled version, it is needed to compute the 
distances from all data points to each prediction point \textit{twice}. The first time is 
carried out to find the $k $ nearest neighbors/data points; and the second is 
to calculate the distance-inverse weights. In this tiled version, both of 
the above two waves of calculating distances are optimized by employing the 
strategy ``tiling''.

By employing the strategy ``tiling'' and exploiting the shared memory, the 
global memory access can be significantly reduced since the coordinates of 
all data points are only read ($n $/threadsPerBlock) times rather than $n$ times 
from global memory, where $n$ is the number of predication points and 
threadsPerBlock denotes the number of threads per block. Furthermore, as 
stated above the strategy ``tiling'' is applied twice.

After calculating each wave of partial distances and weights, each thread 
accumulates the results of all partial weights and all weighted values into 
two registers. Finally, the prediction value of each interpolated point can 
be obtained according to the sums of all partial weights and weighted values 
and then written into global memory.

\section{Results}\label{sec4}
To evaluate the performance of the GPU-accelerated AIDW method, we have 
carried out several groups of experimental tests on a personal laptop 
computer. The computer is featured with an Intel Core i7-4700MQ (2.40GHz) 
CPU, 4.0 GB RAM memory, and a graphics card GeForce GT 730M. All the 
experimental tests are run on OS Windows 7 Professional (64-bit), Visual 
Studio 2010, and CUDA v7.0.

Two versions of the GPU-accelerated AIDW, i.e., the naive version and the 
tiled version, are implemented with the use of both the data layouts SoA and 
AoaS. These GPU implementations are evaluated on both the single precision 
and double precision. However, the CPU version of the AIDW implementation is 
only tested on double precision; and all results of this CPU version are 
employed as the baseline results for comparing computational efficiency. 

All the data points and prediction points are randomly created within a 
square. The numbers of predication points and the data points are 
equivalent. We use the following five groups of data size, i.e., 10K, 50K, 
100K, 500K, and 1000K, where one K represents the number of 1024 (1 K = 
1024). 

For the GPU implementations, the recorded execution time includes the cost 
spent on transferring the input data point from the host to the device and 
transferring the output data from the device back to the host; but it does 
not include time consumed in creating the test data. Similarly, for the CPU 
implementation, the time spent for generating test data is also not 
considered.

\subsection{Single Precision}
\begin{figure}[ht]
	\centering
	\includegraphics[width=0.75\linewidth]{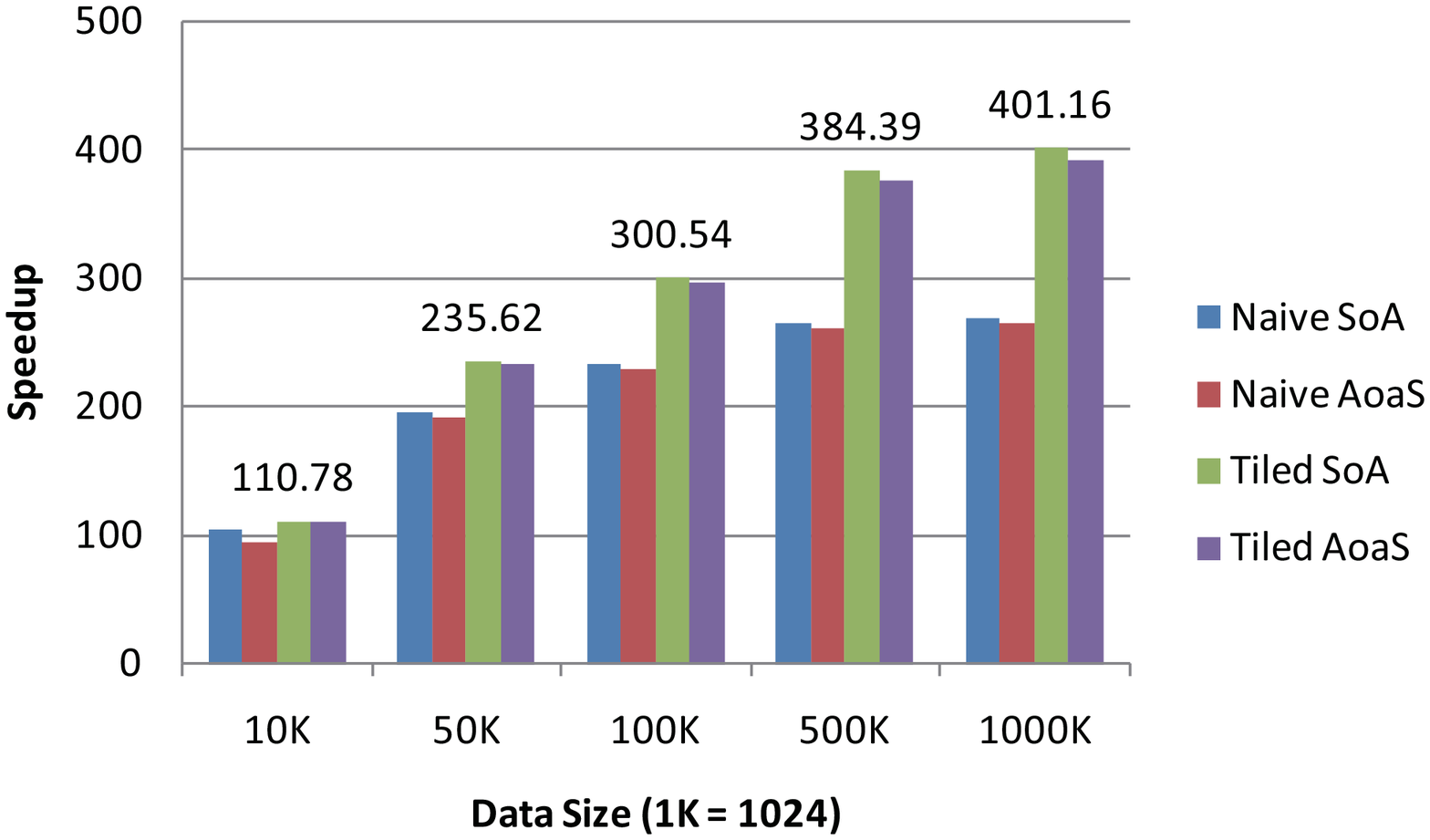}
	\caption{Speedups of the GPU-accelerated AIDW method on single}
	\label{fig4:speedup single}
\end{figure}

The execution time of the CPU and GPU implementations of the AIDW on single 
precision is listed in Table \ref{tab1}. And the speedups of the GPU implementations 
over the baseline CPU implementation are illustrated in Figure \ref{fig4:speedup single}. According 
to these testing results, we have observed that:

(1) The speedup is about 100 $\sim $ 400; and the highest speedup is up to 
400, which is achieved by the tiled version with the use of the data layout 
SoA;

(2) The tiled version is about 1.45 times faster than the naive version;

(3) The data layout SoA is slightly faster than the layout AoaS.

In the experimental test when the number of the data points and 
interpolation points is about 1 million (1000K = 1024000), the execution 
time of the CPU version is more than 18 hours, while in contrast the tiled 
version only needs less than 3 minutes. Thus, to be used in practical 
applications, the tiled version of the GPU-accelerated AIDW method on single 
precision is strongly recommended.

\begin{table}[ht]
	\begin{center}
		\begin{tabular}{|l|l|l|l|l|l|l|}
			\hline
			\raisebox{-1.50ex}[0cm][0cm]{Version}& 
			\raisebox{-1.50ex}[0cm][0cm]{Data  \par Layout}& 
			\multicolumn{5}{|c|}{Data Size (1K = 1024)}  \\
			\cline{3-7} 
			& 
			& 
			10K& 
			50K& 
			100K& 
			500K& 
			1000K \\
			\hline
			CPU& 
			-& 
			6791& 
			168234& 
			673806& 
			16852984& 
			67471402 \\
			\hline
			\raisebox{-1.50ex}[0cm][0cm]{GPU \par Naive}& 
			SoA& 
			65.3& 
			863& 
			2884& 
			63599& 
			250574 \\
			\cline{2-7} 
			& 
			AoaS& 
			66.3& 
			875& 
			2933& 
			64593& 
			254488 \\
			\hline
			\raisebox{-1.50ex}[0cm][0cm]{GPU \par Tiled}& 
			SoA& 
			61.3& 
			714& 
			2242& 
			43843& 
			168189 \\
			\cline{2-7} 
			& 
			AoaS& 
			61.6& 
			722& 
			2276& 
			44891& 
			172605 \\
			\hline
		\end{tabular}
		\caption{\label{tab1}Execution time (/ms) of CPU and GPU implementations of the AIDW method on single precision}
	\end{center}
\end{table}

\subsection{Double Precision}
\begin{figure}[ht]
	\centering
	\includegraphics[width=0.75\linewidth]{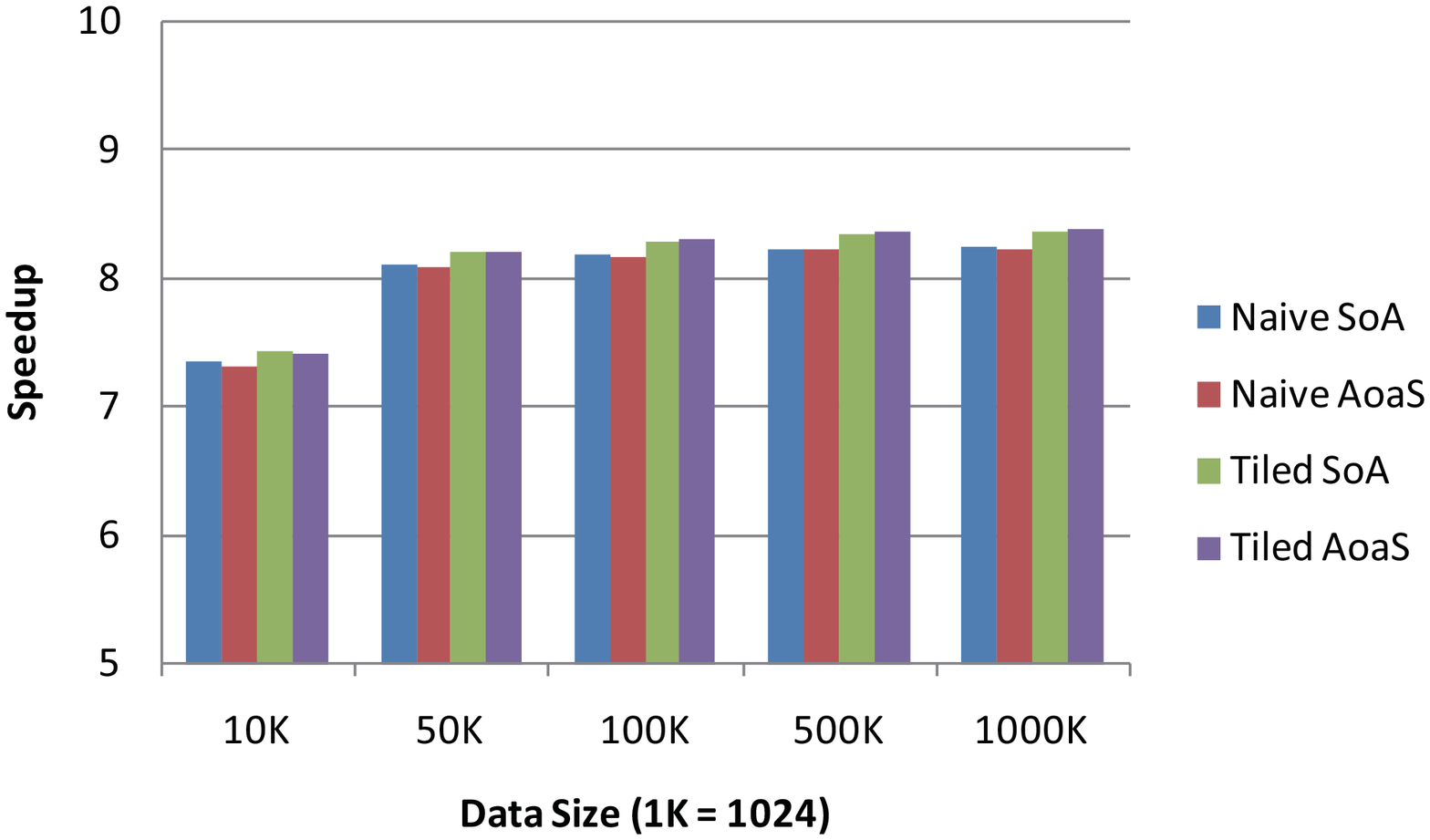}
	\caption{Speedups of the GPU-accelerated AIDW method on double precision}
	\label{fig5:speedup double}
\end{figure}

We also evaluate the computational efficiency of the naive version and the 
tiled version on double precision. It is widely known that the arithmetic 
operated on GPU architecture on double precision is inherently much slower 
than that on single precision. In our experimental tests, we also clearly 
observed this behavior: on double precision, the speedup of the GPU version 
over the CPU version is only about 8 (see Figure \ref{fig5:speedup double}), which is much lower 
than that achieved on single precision. 

We have also observed that: (1) there are no performance gains obtained from 
the tiled version against the naive version; and (2) the use of data 
layouts, i.e., SoA and AoaS, does not lead to significant differences in 
computational efficiency. 

As observed in our experimental tests, on double precision the speedup 
generated in most cases is approximately 8, which means the GPU 
implementations of the AIDW method are far from practical usage. Thus, we 
strongly recommend users to prefer the GPU implementations on single 
precision for practical applications. In the subsequent section, we will 
only discuss the experimental results obtained on single precision.

\section{Discussion}\label{sec5}
\subsection{Impact of Data Layout on the Computational Efficiency}
\begin{figure}[ht]
	\centering
	\includegraphics[width=0.75\linewidth]{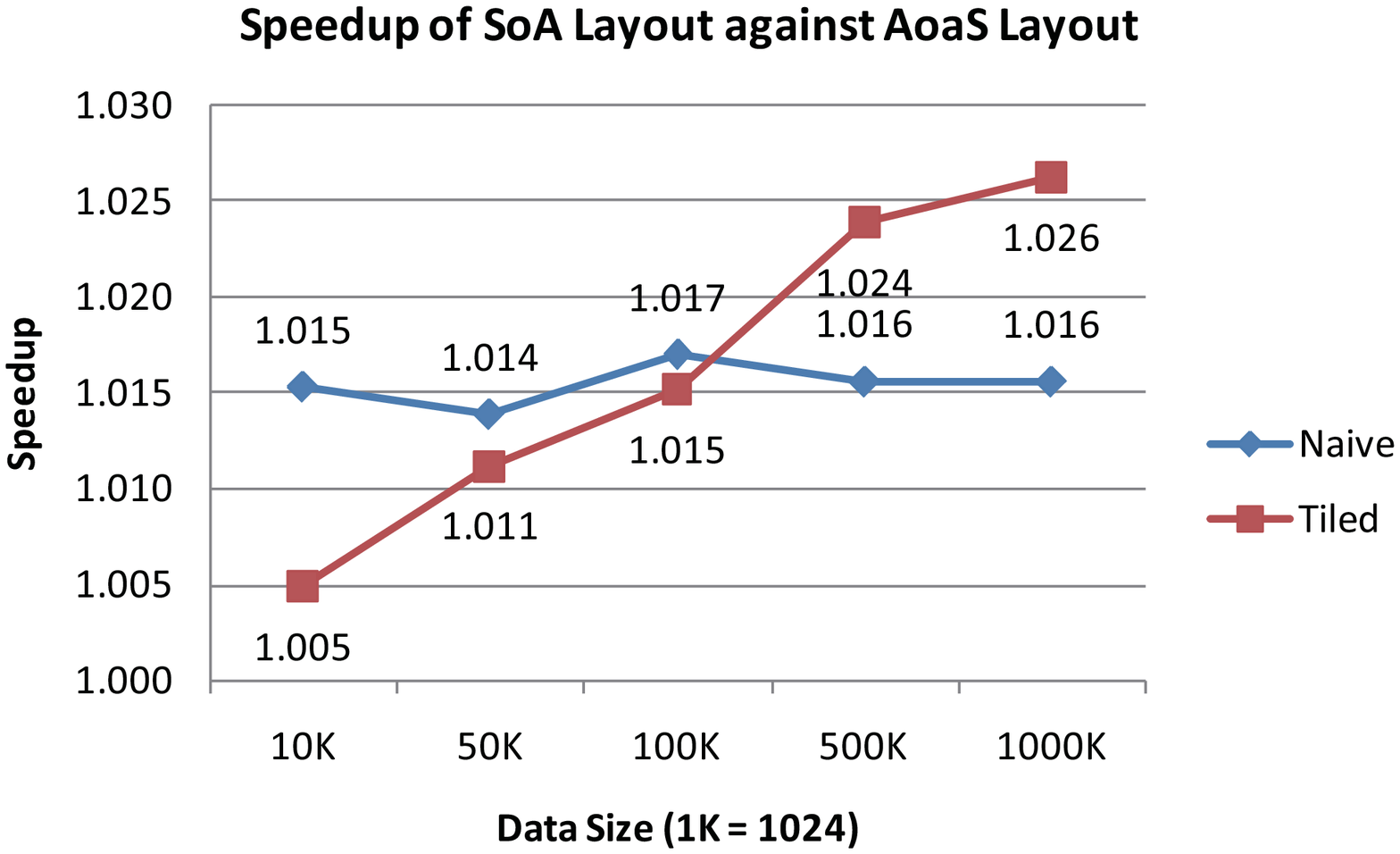}
	\caption{Performance comparison of the layouts SoA and AoaS}
	\label{fig6:soa vs aoas}
\end{figure}

In this work, we have implemented the naive version and the tiled version 
with the use of two data layouts SoA and AoaS. In our experimental tests, we 
have found that the SoA data layout can achieve better efficiency than the 
AoaS. However, there is no significant difference in the efficiency when 
using the above two data layouts. More specifically, the SoA layout is only 
about 1.015 times faster than the AoaS for both the naive version and the 
tiled version; see Figure \ref{fig6:soa vs aoas}.

As stated in Section 3, organizing data in AoaS layout leads to coalescing 
issues as the data are interleaved. In contrast, the organizing of data 
according to the SoA layout can generally make full use of the memory 
bandwidth due to no data interleaving. In addition, global memory accesses 
based upon the SoA layout are always coalesced. This is perhaps the reason 
why the data layout SoA can achieve better performance than the AoaS in our 
experimental tests.

However, it also should be noted that: it is not always obvious which data 
layout will achieve better performance for a specific application. In this 
work, we have observed that the layout SoA is preferred to be used, while in 
contrast the layout AoaS is suggested to be employed in our previous work \cite{31}.

\subsection{Performance Comparison of the Naive Version and Tiled Version}

\begin{figure}[ht]
	\centering
	\includegraphics[width=0.75\linewidth]{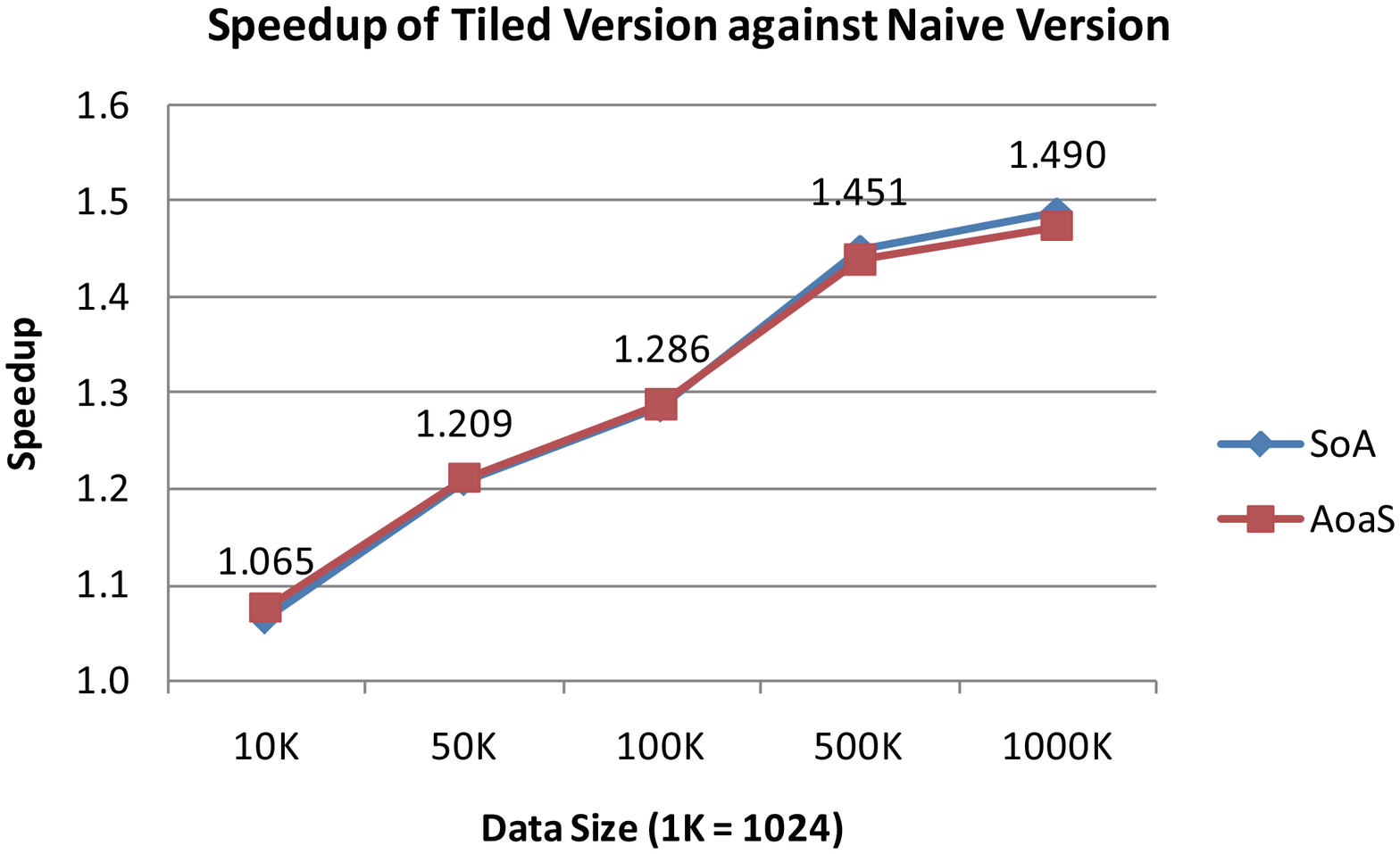}
	\caption{Performance comparison of the naive version and tiled version}
	\label{fig7:naive vs tiled}
\end{figure}

In our experimental tests, we also observed that the tiled version is about 
1.3 times faster than the naive version on average no matter which data 
layout is adopted; see Figure \ref{fig7:naive vs tiled}. This performance gain is due to the use of 
shared memory according to the optimization strategy ``tiling''. 

On GPU architecture, the shared memory is inherently much faster than the 
global memory; thus any opportunity to replace global memory access by 
shared memory access should therefore be exploited. 

In the tiled version, the coordinates of data points originally stored in 
global memory are divided into small pieces/tiles that fit the size of 
shared memory, and then loaded from slow global memory to fast shared 
memory. These coordinates stored in shared memory can be accessed quite fast 
by all threads within a thread block when calculating the distances. By 
blocking the computation this way, we take advantage of fast shared memory 
and significantly reduce the global memory accesses: the coordinates of data 
points are only read ($n$/threadsPerBlock) times from global memory, where 
$n$ is the number of prediction points.

This is the reason why the tiled version is faster than the naive version. 
Therefore, from the perspective of practical usage, we recommend the users 
to adopt the tiled version of the GPU implementations.

\subsection{Comparison of GPU-accelerated AIDW, IDW, and Kriging Methods}

In the literature \cite{32}, Lu and Wong has analyzed the accuracy 
of sequential AIDW with the standard IDW and the Kriging method. They 
focused on illustrating the \textit{effectiveness} of the proposed AIDW method. However, in this 
work, we intend to present a parallel AIDW with the use of the GPU; and here 
we focus on comparing the \textit{efficiency} of the GPU-accelerated AIDW with the 
corresponding parallel version of IDW and Kriging method.

\subsubsection{AIDW vs. IDW}

First, we compare the GPU-accelerated AIDW with the standard IDW. The AIDW 
is obviously more computationally expensive than the IDW. The root cause of 
this behavior is also quite clear: in AIDW it is needed to \textit{dynamically} determine the 
adaptive power parameter according to the spatial points' distribution 
pattern during interpolating, while in IDW the power parameter is 
\textit{statically} set before interpolating. 

Due to the above reason, in the GPU-accelerated AIDW, each thread needs to 
perform much more computation than that in the IDW. One of the obvious extra 
computational steps in the GPU-accelerated AIDW is to find some nearest 
neighbors, which needs to loop over all the data points. This means in the 
AIDW each thread needs to calculate the distances from all the data points 
to one interpolated point twice: the first time is to find the nearest 
neighbors and the second is to obtain the distance-inverse weights. In 
contrast, in the IDW each thread is only invoked to calculate the distances 
once.

\subsubsection{AIDW vs. Kriging}

Second, we compare the GPU-accelerated AIDW with the Kriging method. 
Although the AIDW is by nature computationally expensive than the IDW, it is 
still computationally inexpensive than the Kriging method. In Kriging, the 
desired prediction value of each interpolated point is also the weighted 
average of data points. The essential difference between the Kriging 
interpolation and the IDW and AIDW is that: in Kriging the weights are 
calculated by solving a linear system of equations, while in both IDW and 
AIDW the weights are computed according to a function in which the parameter 
is the distance between points. 

As stated above, in Kriging the weights are achieved by solving a linear 
system of equations, $AV = B$, where $A$ is the matrix of the semivariance 
between points, $V$ is the matrix of weights and $B$ stands for the variogram. 
Thus, in fact, the Kriging method can be divided into three main steps: (1) 
the assembly of $A$, (2) the solving of $AV = B$, and (3) the weighting average. 
The third step is the same as those in both IDW and AIDW. The most 
computationally intensive steps are the assembly of $A$ and the solving of $AV = B$.

Each entry of the matrix $A$ is calculated according to the so-called 
semivariance function. The input parameter of the semivariance function is 
the distance between data points. Thus, in GPU-accelerated Kriging method, 
when each thread is invoked to compute each row of the matrix $A$, it is 
needed to loop over all data points for each thread to first calculate the 
distances and then the semivariances. Obviously, the assembly of the matrix 
A can be performed in parallel by allocating $m$ threads, where $m$ is the number 
of data points.

Another key issue in assembling the matrix $A$ is the storage. In AIDW, the 
coordinates of all the data points and interpolated points are needed to be 
stored in global memory. There is no other demand for storing the large size 
of intermediate data during interpolating. However, in Kriging the matrix $A$ 
needs to be additionally stored during interpolating. The matrix $A$ is a 
square matrix with the size of \textit{at least} ($m$+1)$\times $ ($m$+1), where $m$ is the number of 
data points. 

Moreover, in general the $A$ is dense rather than sparse, which means the 
matrix $A$ cannot be stored in any compressed formats such as COO (COOrdinate) 
or CSR (Compressed Sparse Row). Thus, the global memory accesses in Kriging 
are much more than those in AIDW due to the storing of the matrix $A$. This 
will definitely increase the computational cost. 

As mentioned above, the vector of weights V can be obtained by solving the 
equations $AV = B$. But in fact the vector can be calculated according to 
$V=A^{-1}B$. This means the matrix $A$ is only needed to be inverted once, and 
then repeatedly used. However, to obtain the vector of weights V, first the 
inverse of the matrix $A$ and then the matrix multiplication is needed to be 
carried out. The matrix inverse is only needed to be performed once; but the 
matrix multiplication has to be carried out for each interpolated point.

In GPU-accelerated Kriging method, when a thread is invoked to predict the 
value of one interpolated point, then each thread needs to perform the 
matrix multiplication $V=A^{-1}B$. The classical matrix multiplication 
algorithm is computationally expensive in CPU as a consequence of having an 
$O(n^{3})$ complexity. And, the complexity of the matrix multiplication 
algorithm in a GPU is still $O(n^{2})$ \cite{18}. This means 
in GPU-accelerated Kriging method, each thread needs to perform the 
computationally expensive step of matrix multiplication. In contrast, in 
GPU-accelerated AIDW, the most computationally expensive step is the finding 
of nearest neighbors, which only has the complexity of $O(n)$.

\section{Conclusions}\label{sec6}
On a single GPU, we have developed two versions of the GPU-accelerated AIDW 
interpolation algorithm, the naive version that does not profit from shared 
memory and the tiled version that takes advantage of shared memory. We have 
also implemented the naive version and the tiled version with the use of two 
data layouts, AoS and AoaS, on both single precision and double precision. 
We have demonstrated that the naive version and the tiled version can 
approximately achieve the speedups of about 270 and 400 on single precision, 
respectively. In addition, on single precision the implementations using the 
layout SoA are always slightly faster than those using the layout AoaS. 
However, on double precision, the overall speedup is only about 8; and we 
have also observed that: (1) there are no performance gains obtained from 
the tiled version against the naive version; and (2) the use of data 
layouts, i.e., SoA and AoaS, does not lead to significant differences in 
computational efficiency. Therefore, the tiled version that is developed 
using the layout SoA on single precision is strongly recommended to be used 
in practical applications.

\section*{Acknowledgements}
This research was supported by the Natural Science Foundation of China 
(Grant No. 40602037 and 40872183), China Postdoctoral Science Foundation 
(2015M571081), and the Fundamental Research Funds for the Central 
Universities (2652015065).

\bibliography{myref}

\end{document}